%% file: gecco.tex
\documentclass{sig-alternate}

\usepackage{algorithmic}
\usepackage{algorithm}
\usepackage{amsmath}

\usepackage{graphicx}
\usepackage{mathtools}
\usepackage[table]{xcolor}

\newtheorem{RedRule}{Reduction Rule}
\newtheorem{definition}{Definition}

\begin{document}

%\conferenceinfo{GECCO'13,} {July 6-10, 2013, Amsterdam, The Netherlands.}
%\CopyrightYear{2013}
%\crdata{TBA}
%\clubpenalty=10000
%\widowpenalty = 10000

\title{Improving CASA Runtime Performance by Exploiting Basic Feature Model Analysis}%Think of a new title, see if we can manage to mention {\ttlit simulated annealing} in the title \titlenote{This might be good as we hope to get into search based track of GECCO}}

\numberofauthors{3}
%\author{
%\alignauthor Evelyn Nicole Haslinger\\
% \affaddr{Institute for Systems Engineering and Automation}\\
% \affaddr{Johannes Kepler University}\\
% \affaddr{Linz, Austria}\\
% \email{evelyn.haslinger@jku.at}
%\and
%\alignauthor Roberto E. Lopez-Herrejon\\
% \affaddr{Institute for Systems Engineering and Automation}\\
% \affaddr{Johannes Kepler University}\\
% \affaddr{Linz, Austria}\\
% \email{roberto.lopez@jku.at}
%\alignauthor Alexander Egyed\\
% \affaddr{Institute for Systems Engineering and Automation}\\
% \affaddr{Johannes Kepler University}\\
% \affaddr{Linz, Austria}\\
% \email{alexander.egyed@jku.at}  
\author{      
\alignauthor  Evelyn Nicole Haslinger\\
 \affaddr{Institute for Systems Engineering and Automation}\\
 \affaddr{Johannes Kepler University}\\
 \affaddr{Linz, Austria}\\
 \email{evelyn.haslinger@jku.at}   
\alignauthor  Roberto E. Lopez-Herrejon\\
 \affaddr{Institute for Systems Engineering and Automation}\\
 \affaddr{Johannes Kepler University}\\
 \affaddr{Linz, Austria}\\
 \email{roberto.lopez@jku.at}   
  \alignauthor  Alexander Egyed\\
 \affaddr{Institute for Systems Engineering and Automation}\\
 \affaddr{Johannes Kepler University}\\
 \affaddr{Linz, Austria}\\
 \email{alexander.egyed@jku.at}              
}
\maketitle

\begin{abstract}
In Software Product Line Engineering (SPLE) families of systems are designed, rather than developing the individual systems independently. 
Combinatorial Interaction Testing has proven to be effective for testing in the context of SPLE, where a representative subset of products is chosen for testing in place of the complete family. 
Such a subset of products can be determined by computing a so called t-wise Covering Array (tCA), whose computation is NP-complete. 
%We strongly believe that the generation of tCAs in the context of SPLE is extremely well suited for search based techniques. 
Recently,  reduction rules that exploit basic feature model analysis have been proposed that reduce the number of elements that need to be considered during the computation of tCAs for Software Product Lines (SPLs). We applied these rules to CASA, a simulated annealing algorithm for tCA generation for SPLs. We evaluated the adapted version of CASA using 133 publicly available feature models and could record on average a speedup of 61.8\% of median execution time, while at the same time preserving the coverage of the generated array.  
%Moreover we performed a thorough statistical analysis of the obtained results to ensure that our observed improvements are not due to the non-deterministic nature of CASA. 

%Combinatorial Interaction Testing has shown great potential for effectively testing Software Product Lines (SPLs).
%An important part of this type of testing is determining a subset of SPL products in which interaction errors are more likely to occur.
%Such sets of products are obtained by computing a so called t-wise Covering Array (tCA), whose computation is known to be NP-complete. 
%Recently, the ICPL algorithm has been proposed to compute these covering arrays.
%In this research-in-progress paper, we propose a set of rules that exploit basic feature model knowledge to reduce the number of elements (i.e. t-sets) 
%required by ICPL without weakening the strength of the generated arrays.
%We carried out a comparison of runtime performance that shows a significant reduction of the needed execution time for the majority of our SPL case studies. 
\end{abstract}

%A category including the fourth, optional field follows...



\input{Introduction.tex}

\input{Background.tex}

\input{ReductionRules.tex}

\input{SpeedingUpCasa.tex}

\input{Evaluation.tex}
\input{RelatedWork.tex}
\input{ConclusionsFutureWork.tex}

%\section*{Acknowledgment}

\section*{Acknowledgment}
This research is partially funded by the Austrian Science Fund (FWF) project P21321-N15 and Lise Meitner Fellowship M1421-N15.
%Ugaitz research was sponsored by the
%Erasmus exchange program and the government of the Basque Country
%Spain. Currently, he is partially sponsored by the Universitat Politecnica
%de Catalunya.

\bibliographystyle{abbrv}
\bibliography{biblio}  % sigproc.bib is the name of the Bibliography in this case
% ACM needs 'a single self-contained file'!
%
\end{document}

%% file: Introduction.tex
\section{Introduction} \label{sec:Intro}
Because of changes in user requirements, or due to varying requirements among the companies' customers, software products are less and less frequently designed as one-of-a-kind systems. 
In the \textit{Software Product Line Engineering (SPLE)} paradigm the emphasis lies on designing families of systems rather than designing the individual systems separately \cite{CE00, SPLE, BSR04}. 
The core of SPLE is a \textit{Software Product Line (SPL)}, where an SPL describes a family of software products that usually offer similar functionality \cite{CE00, PBL05, BSR04}. The member products of an SPL can be distinguished by the set of features they implement, where a \textit{feature} is often defined as an increment in program functionality \cite{zave}. \textit{Feature Models (FMs)} are considered the de-facto standard to model the commonalities, differences and relationships among the features implemented by the member products of an SPL \cite{CE00, KCH+90}.

Testing is a vital and often time intensive task in any software project's life cycle. In SPLE a family of systems needs to be tested rather than a single system. 
%As in other research areas SPLE imposes, so to speak, another dimension of complexity to the task at hand, because rather than to design tests that show that a single system conforms to its specification it needs to be shown that \texttt{n} systems conform to their specifications. 
In other words, SPLE imposes another dimension of complexity to the testing task: namely, rather than designing tests to show that a single system conforms to its specification the tests should be designed to show that the \texttt{n} systems conform to the SPL specifications. 
While of course single system testing techniques could be applied to the individual member products of the SPL, this may be rather ineffective as the products of an SPL usually share a considerable amount of source code. 
Different approaches have been proposed to overcome some of the difficulties of testing SPLs \cite{LKL12, ER11, NL11, RRKP06, UKB10}. 

One SPL testing approach is Combinatorial Interaction Testing, whereby a subset of products is chosen to be tested in such a way that interaction errors are the most likely to occur \cite{Cohen08}. 
In this work we focus on Combinatorial Interaction Testing, or more precisely on the selection of a proper subset of products (i.e. so called \textsl{t-wise Covering Arrays (tCAs)}) for testing SPLs. 
Garvin et. al proposed CASA \cite{GCD11}, a simulated annealing algorithm, for the generation of tCAs for SPLs. 
Earlier work by Haslinger et. al proposes reduction rules to reduce the number of elements that need to be considered during the generation of tCAs \cite{VaMosPaper13}. In this paper, we combine these two works by showing how these reduction rules can be applied to CASA. 
We evaluated our new version of CASA using 133 publicly available FMs from the SPLOT website \cite{SPLOT}. We run the original CASA and its adapted version each a hundred times per model and recorded on average a speed-up of 61.8\% of median execution time. 
Moreover we performed a thorough statistical analysis, that shows that our adaptations indeed result in shorter runtimes and are not simply due to the non-deterministic nature of CASA. 
Hence, we provide further proof that the proposed reduction rules lead to significantly shorter runtimes when they are applied to tCA generation algorithms. 
Additionally, we compared the generated arrays showing that our reduction rules do not have any effect on the size of the generated arrays. In other words, the adapted version of CASA generates the tCAs faster while still preserving their strength.

%% file: Background.tex
\section{Background}
This section presents the basic concepts on variability modeling with feature models and combinatorial interaction testing.% and simulated annealing. 

\subsection{Feature Models}
The member products of an SPL are distinguished by the set of features they implement. Usually there exist constraints among the features in the SPL that each member product needs to adhere to. For instance, two features may mutually exclude each other or the selection of one feature might necessitate the selection of another one. The de-facto standard to model such constraints among features in SPLE are \textit{Feature Models (FMs)} \cite{CE00, KCH+90}. 

\begin{figure}
\centering{}\includegraphics[scale=0.26]{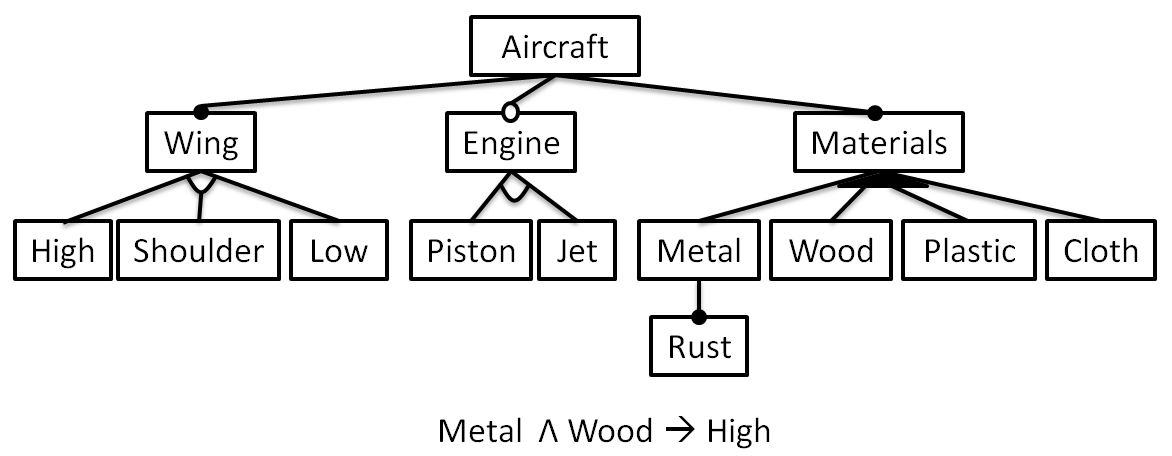}\caption{Feature Model of Aircraft SPL}
\label{fig:FM} 
\end{figure}

FMs are tree-like structures, where each node represents a feature of the SPL. An example of an FM for an Aircraft SPL is shown in Figure~\ref{fig:FM}\footnote{Note that this FM is based on an FM that is part of the corpus we used to evaluate our work \cite{SPLOT}.}. 
Each FM has a single root feature that is implemented by each member product of the SPL. Feature \texttt{Aircraft} is the root feature of our running example, meaning that each valid product of the Aircraft SPL needs to include feature \texttt{Aircraft}. 

Each feature, apart from root, has a single parent feature with which it can interrelate in four different kinds of relationships. \textit{Mandatory} child features need to be included whenever their parent feature is included, these features are denoted using filled circles. For instance feature \texttt{Wing} is a mandatory child of feature \texttt{Aircraft}. For \textit{optional} child features holds that they may or may not be selected whenever their parent is included, this kind of relationship is depicted by empty circles. An example of an optional feature in our running example is feature \texttt{Engine}. \textit{Inclusive-or} groups are denoted by filled arcs, for these features holds that at least one feature of the group needs to be selected whenever their parent is selected. In the Aircraft FM features \texttt{Metal}, \texttt{Wood}, \texttt{Plastic} and \texttt{Cloth} constitute an inclusive-or. The last type of child parent relation is the \textit{exclusive-or} group, which is depicted using an empty arc. For an exclusive-or group holds that exactly one feature of the group needs to be selected whenever their parent is included. In our running example features \texttt{Piston} and \texttt{Jet} form an exclusive-or. 

Apart from these child parent relationships there are also so called \textit{Cross Tree Constraints} (CTCs). CTCs express arbitrary relationships among the features within the SPL and are usually denoted using propositional logic formulas \cite{Benavides2010}. An example of a CTC is \texttt{Metal}~$\wedge$~\texttt{Wood} $\Rightarrow$~\texttt{High}, which means that whenever feature \texttt{Metal} and feature \texttt{Wood} are selected feature \texttt{High} must also be selected.

\subsection{Combinatorial Interaction Testing}

\textit{Combinatorial Interaction Testing (CIT)} is an approach to tackle some of the issues imposed by testing SPLs, i.e. the large number of products that need to be tested and the redundant test effort that results from the source code that is common to several products of the SPL. When CIT is to be applied there are essentially three consecutive steps that need to take place: 
\begin{enumerate}
	\item Choose a representative subset of products for testing where interaction faults are likely to occur. 
	\item Configure and implement the products of this subset.
	\item Apply single system testing techniques on these products. 
\end{enumerate}

In this paper we focus on the first step of CIT, i.e. on selecting a proper subset of products for testing. One possibility to choose this subset are \textit{t-wise covering arrays (tCAs)}, that we shall describe next. First, let us introduce some basic terminology. As mentioned before, the products within an SPL can be distinguished by the set of features they implement, more formally they can be represented using feature sets \cite{VaMosPaper13, FasePaper13}. 

\begin{definition} \textit{Feature List (FL) is the list of features
in a software product line.}
\end{definition}

\begin{sloppypar}
The feature list of our running example is \texttt{FL = \{Aircraft, Wing, Engine, Materials, High, Shoulder, Low, Piston, Jet, Metal, Wood, Plastic, Cloth, Rust\}}.
\end{sloppypar}

\begin{definition} \textit{Feature Set (FS) is a 2-tuple {[}sel,$\overline{sel}${]}
where sel and $\overline{sel}$ are respectively the set of selected
and not-selected features of a member product. Let FL be a feature
list, thus sel, $\overline{sel}$ $\subseteq$ FL, sel $\cap$ $\overline{sel}$
= $\emptyset$, and sel $\cup$ $\overline{sel}$ = FL. The terms
p.sel and p.$\overline{sel}$ respectively refer to the set of selected
and not-selected features of product p\footnote{Definition based on \cite{Benavides2010}.}.}
\end{definition}

An example of a feature set for our running example is: 

\begin{equation} \label{eq:ts1}
\begin{multlined}
\texttt{fs}_1=[\{\texttt{Aircraft}, \texttt{ Wing}, \texttt{ High}, \texttt{ Materials}, \texttt{ Metal}, \\
 \shoveleft[0.5cm] \texttt{ Rust}\}, \{\texttt{ Engine}, \texttt{ Shoulder}, \texttt{ Low}, \texttt{ Piston}, \texttt{ Jet}, \\
 \shoveleft[0.13cm] \texttt{ Wood},  \texttt{ Plastic}, \texttt{ Cloth}\}]
\end{multlined}
\end{equation}

\begin{definition} \textit{A t-set ts is a is a 2-tuple {[}sel,$\overline{sel}${]} representing a partially configured product, defining the selection of t features of feature list FL, i.e. $ts.sel \cup ts.\overline{sel} \subseteq FL $ $\land$ $ts.sel \cap ts.\overline{sel}=\emptyset$ $\land$ $|ts.sel \cup ts.\overline{sel}| = t $. We say t-set ts is covered by feature set fs iff $ts.sel \subseteq fs.sel $ $\land$ $ ts.\overline{sel} \subseteq fs.\overline{sel}$.}
\end{definition}  

\begin{definition} \textit{We say a feature set \texttt{fs} is valid in feature model \texttt{fm}, i.e. \texttt{valid(fs, fm)} holds, iff \texttt{fs} does not contradict any of the constraints introduced by \texttt{fm}\footnote{Such constraints are introduced by child parent relationships and additional CTC that are expressed as propositional logic formulas and are obtained as described in \cite{Benavides2010}.}. A t-set \texttt{ts} is valid if there exists a valid feature set \texttt{fs} that covers \texttt{ts}. 
}
\end{definition}

Feature set \texttt{fs$_1$} shown in Equation~\ref{eq:ts1} is valid with respect to the feature model of the Aircraft SPL. An example of a valid 3-set would be \texttt{[\{Wing, Piston\}, \{Wood\}]}. The 3-set \texttt{[\{Metal, Wood\}, \{High\}]} is invalid with respect to the feature model of our running example, because it contradicts the CTC \texttt{Metal $\land$ Wood $\Rightarrow$ High}, which states that feature \texttt{High} must also be selected. 

A t-wise CA can be formally defined as: 
\begin{definition} \textit{A t-wise covering array tCA for a feature model FM is a set of valid feature sets that covers all valid t-sets of the software product line\footnote{Definition inspired by \cite{JHF12}.}}.
\end{definition}

Kuhn et. al have shown in \cite{KWG04} that when a tCA of strength t=1 is used, then it is likely that 50\% of the defects are found. When a 2-wise CA is used then it is likely to find 70\% of the defects. If t is increased to 3, i.e. a 3-wise CA is used for the test suit genereation, then it is likely that even 95\% of the defects are found.
%\begin{enumerate}
%	\item When a tCA of strength 1 is used, it is likely that  
%	\item When a tCA of strength 2 is used, it is likely that 70\% of the defects are found. 
%	\item When a tCA of strength 3 is used, it is likely that 95\% of the defects are found. 
%\end{enumerate}

For a 1-wise CA holds that each feature is in at least one feature set of the 1-wise CA selected respectively deselected. Meaning we make sure to test whether a product behaves faulty when a certain feature is included or not included. For a 2-wise CA holds that each combination among two features is covered by at least one of the feature sets in the 2-wise CA, meaning that all four possible interactions among two features \texttt{f$_1$} and \texttt{f$_2$} are covered by the chosen set of products.

The generation of tCAs is an NP complete problem\cite{JHF12, VaMosPaper13}. It is common knowledge that there are still no algorithms known that are guaranteed to find optimal solutions to problems in this complexity class in an efficient way. This fact makes the generation of tCAs for constrained problems within the context of SPLE extremely well suited for SBSE techniques.

%% file: ReductionRules.tex
\section{Reduction Rules for t-sets}

Apart from applying a search based algorithm to the generation of tCAs it is also vital to think about possibilities to reduce the search space of the problem. This section outlines two reduction rules for t-sets based on an earlier work of Haslinger et. al \cite{VaMosPaper13} that can be used to reduce the number of t-sets that need to be considered during the generation of a tCA. 

Any algorithm that computes a t-wise CA for a feature model \texttt{fm} needs to determine two sets:

\begin{itemize}
	\item The set of valid t-sets \texttt{V} in \texttt{fm}.
	\item A set \texttt{tCA} of valid feature sets that covers all t-sets in \texttt{V}.   
\end{itemize}

There exists a subset \texttt{V'} of \texttt{V} (i.e. \texttt{V' $\subseteq$ V}) for which holds that if \texttt{tCA} covers all t-sets in \texttt{V'} then it automatically covers all elements in \texttt{V}. In the remainder of this section we outline two reduction rules that can be used to compute \texttt{V'} \cite{VaMosPaper13}.  

\begin{RedRule} Any t-set t that contains the root feature r of the feature model (i.e. $ r \in t.sel \cup t.\overline{sel}$) is neglectible during the computation of a tCA. 
\end{RedRule}

Consider for instance the t-set \texttt{t$_1$=[\{Aircraft\}, \{Engine, Wood\}]} for the feature model of our running example. According to Reduction Rule 1 we do not need to cover this t-set. The t-set \texttt{t$_2$=[\{Metal\}, \{Engine, Wood\}]} is an element of \texttt{V'} as none of the presented reduction rules applies. A tCA, which contains a feature set \texttt{fs} that covers \texttt{t$_2$} automatically also covers  \texttt{t$_1$}. If \texttt{fs} covers \texttt{t$_2$} then \texttt{\{Engine, Wood\} $\subseteq$ fs.$\overline{\texttt{sel}}$} and by definition any valid feature set needs to select the root feature, i.e. \texttt{\{Aircraft\} $\subseteq$ fs.sel}, hence \texttt{fs} covers also \texttt{t$_1$}.

\begin{RedRule} Any t-set t that contains a mandatory child feature m of the feature model (i.e. $ m \in t.sel \cup t.\overline{sel}$) is neglectible during the computation of a tCA. 
\end{RedRule}

\begin{sloppypar}
Consider the t-set \texttt{t$_1$=[\{Rust, Shoulder\},\{Plastic\}]} and t-set \texttt{t$_2$=[\{Metal, Shoulder\},\{Plastic\}]} once again the proposition holds that  \texttt{t$_2$} is an element of \texttt{V'} as none of the reduction rules applies to this t-set. A valid feature set \texttt{fs} covering \texttt{t$_2$} also covers \texttt{t$_1$}, because by definition \texttt{Rust} needs to be selected whenever \texttt{Metal} is included hence \texttt{\{Rust, Shoulder\} $\subseteq$ fs.sel} and \texttt{\{Plastic\} $\subseteq$ fs.$\overline{\texttt{sel}}$}.
\end{sloppypar}

Note that these reduction rules have two advantages, do not only reduce the set of valid t-sets \texttt{V} but can be used to generate a smaller set of possible t-sets, i.e. for a feature model with feature list \texttt{FL} there are  $\frac{|FL|!}{t!*(|FL|-t)!} * 2^t $ different t-sets that all need to be checked whether they are valid in the considered feature model\footnote{The expression $\frac{|FL|!}{t!*(|FL|-t)!}$ denotes the number of permutations containing \texttt{t} elements from a set of size \texttt{|FL|}, and $2^t$ represents the number of possible combinations among t features \cite{VaMosPaper13}.}. Let \texttt{reduceable} denote the set containing the root feature and all mandatory child features, i.e. all features that are neglectible according to the presented reduction rules. When the proposed reduction rules are applied only $\frac{|FL \setminus reduce able|!}{t!*(|FL \setminus reduce able|-t)!} * 2^t $ t-sets need to be checked for validity. 

Please refer to~\cite{VaMosPaper13} for a more formal proof of the two presented reduction rules.

%% file: SpeedingUpCasa.tex
\section{Speeding Up CASA} \label{sec:CASA}
In this paper we focus on CASA a meta-heuristic search using simulated annealing to generate t-wise CAs for constrained problems \cite{GCD11}. To the best of our knowledge it is the only search based approach to generate t-wise CAs in the context of SPLE, this being the main reason why we chose CASA to evaluate the proposed reduction rules. In the remainder of this section we will outline how we adapted CASA to incorporate the proposed reduction rules. 

\subsection{Data Structures for Input Models}
The Orthogonal Variability Modeling Language (OVM) is an alternative to FMs in SPLE \cite{SPLE}. CASA takes two input files that are used to decode an OVM model. 

One specifies the properties of the array to generate, namely its strength, the number of columns and the valid values per column. This information is decoded using three lines, the first specifies the strength \texttt{t} of the array, the second defines the number of features and the third specifies the number of possible values a feature can have. An input file defining the properties of the array for our running example would be for example:

3

14

2 2 2 2 2 2 2 2 2 2 2 2 2 2

Which means that CASA is instructed to generate a tCA of strength 3, with 14 columns, where each column can take take 2 different values. Note here, that as we are considering FMs the only valid number of possible values is 2, i.e. either we select or deselect a feature \texttt{f}. The first column of the array to generate can take values 0 or 1, the second column can take the values 2 or 3 and so forth. If feature \texttt{f} represented by column \texttt{n} is selected (resp. deselected) we denote this by the value $2*(n-1)$ (resp. $2*(n-1)+1$). Consider the first row of Table~\ref{tab:Mappings}. It shows a possible encoding for the features of our running example. For instance, feature \texttt{Low} is decoded with value 12 when selected and with value 13 when deselected. The feature set shown in Equation~\ref{eq:ts1} would be encoded by the following row [0, 2, 5, 6, 8, 11, 13, 15, 17, 18, 19, 21, 23, 25, 26]. 

The second input file defines the constraints that specify a valid row in the array to generate, i.e. it denotes the constraints imposed by the tree-structure of the FM as well as its CTCs. These constraints are denoted in CNF over the valid values per row, which range from $0$ to $2*(n-1)+1$. If we want for instance to denote that feature \texttt{f$_1$} (represented by the first columns) requires the selection of a feature \texttt{f$_2$} (represented by the second column) we would denote this using the proposition $\neg 0 \lor 2$\footnote{Note that $1 \lor 2$ would be equivalent.}. To denote that feature \texttt{Piston} mutually excludes feature \texttt{Jet} we would write $\neg 14 \lor \neg 16$, according to the encoding shown in the first row of Table~\ref{tab:Mappings}. The CTC of our running example \texttt{Metal}~$\wedge$~\texttt{Wood} $\Rightarrow$~\texttt{High} which is equivalent to $\neg$\texttt{Metal}~$\lor$~$\neg$\texttt{Wood} $\lor$~\texttt{High} would be encoded as $\neg 18 \lor \neg 20 \lor 8$ or alternatively $19 \lor 21 \lor 8$.

The first necessary step of our evaluation was to convert our FMs to this format. To do so, we used the ICPL tool provided by the resources homepage of Johansen et al. \cite{JHF12}. 

\subsection{Adaptations to CASA}
The next step was to actually enhance the provided CASA implementation with the proposed reduction rules. It turns out that CASA has been designed in a way that does not permit changes in the search without the need to restructure various parts of the source code. So we chose to adapt the constraints and number of features before they are passed to the annealing procedure and to adapt the generated array before it is written to the output file. 

\begin{algorithm}                      
% enter the algorithm environment 
\caption{Adapted Overall Procedure of CASA}
\label{al:overall}         
% give the algorithm a caption \label{alg1}                           
% and a label for \ref{} commands later in the document 
\begin{algorithmic}[1]                    
% enter the algorithmic environment 
%\REQUIRE $n \geq 0 \vee x \neq 0$ 
%\ENSURE $y = x^n$
\STATE Input: Path to file defining the array properties (prop) and a path to file containing the constraints among features (constr)
% of requiring type $[F, e, TS, FC, requiring]$ with $FC \neq \emptyset$$_{conf}$, and $PL_{f}$.
\STATE Output: A t-wise CA
\STATE
%\STATE
\STATE \COMMENT{Read in the features and the constraints among them}
\STATE $features := readProperties(prop)$ \label{al:readProp}
\STATE $constrs := readConstraints(constrs)$ \label{al:readConstrs}
\STATE
\STATE \COMMENT{Generate Mapping from old to new feature values} \label{al:startAdapt}
\STATE $(root, mandatories) := $\\  \quad \quad $findMandAndRoot(features, constrs)$ \label{al:getMandRoot}
\STATE $(oldToNew, newToOld) := $\\  \quad \quad $generateMappings(root, mandatories)$ \label{al:getMapping}
\STATE
\STATE \COMMENT{Reduce considered features and constraints}
\STATE $features' := reduceFeatures(root, mandatories)$ \label{al:redFeatures}
\STATE $constrs' := adaptConstrs(oldToNew)$ \label{al:endAdapt}
\STATE
\STATE \COMMENT{Call CASA's annealing procedure}
\STATE $tCA = anneal(features', constrs')$ \label{al:anneal}
\STATE
\STATE \COMMENT{Build Feature Model}
\STATE $tCA' = expand(tCA, newToOld, mandatories)$ \label{al:expand}
\STATE
\RETURN $tCA'$ \label{al:return}
\end{algorithmic} 
\end{algorithm}

Let us first outline the basic idea of this adaptation. The CASA implementation assumes that the possible values for the individual columns in the array to generate reach from $0$ to $2*\texttt{n}-1$, where \texttt{n} denotes the number of columns. For our running example, the valid values for the individual columns reach from $0$ to $2*14-1=27$, as we have 14 features in our input FM.
If the input model contains \texttt{m} features that can be neglected according to the reduction rules, then the annealing procedure of CASA now works on values from $0$ to $2*(\texttt{n}-\texttt{m})-1$. As there are 4 features in our running example that are either a mandatory child or the root feature, CASA now works on values ranging from 0 to 19. A feature that was originally denoted by column \texttt{m}, is now denoted by column $\texttt{m'} \leq \texttt{m}$, which clearly necessitates the adaptation of the input constraints. In order to adapt the input constraints accordingly we need to come up with a mapping from old to new values. The annealing procedure of CASA now generates an array that corresponds to the adapted constraints and value range, we therefore also need a mapping from new values to old values in order to post process the generated array before it is written to the output file.

Consider Algorithm~\ref{al:overall}. It depicts the adapted overall procedure of CASA. Essentially only Lines~\ref{al:startAdapt}~to~\ref{al:endAdapt} and Line~\ref{al:expand} were added for our evaluation. First CASA reads in the features of the SPL, i.e. the properties of the array to generate that are specified in the file \texttt{prop} (see Line~\ref{al:readProp}). Next the constraints among these features are read in and stored in variable \texttt{constraints} (see Line~\ref{al:readConstrs}). 

%\scalebox{0.7}{
\begin{table*} 
	\centering
	\scriptsize
		\begin{tabular}{|c|c|c|c|c|c|c|c|c|c|c|c|c|c|c|}%{|p{15mm}|p{5mm}|p{5mm}|p{12mm}|p{13mm}|p{10mm}|}
			\hline 
			 &\textbf{Airc.}&\textbf{Wing}&\textbf{Eng.} &\textbf{Mat.}&\textbf{High}&\textbf{Shoul.}&\textbf{Low}&\textbf{Piston}&\textbf{Jet}&\textbf{Metal}&\textbf{Wood}&\textbf{Plastic}&\textbf{Cloth}&\textbf{Rust}\tabularnewline
			\hline 
			\hline 
			\textbf{old} & 0, 1 & 2, 3 & 4, 5 & 6, 7 & 8, 9 & 10, 11 & 12, 13 & 14, 15 & 16, 17 & 18, 19 & 20, 21 & 22, 23 & 24, 25 & 26, 27  \tabularnewline
		\hline 
		\hline
	\textbf{new} &\cellcolor{gray!20} 20 , 21 &\cellcolor{gray!20} 20, 21 & 0, 1 &\cellcolor{gray!20} 20, 21 & 2, 3 & 4, 5 & 6, 7 & 8, 9 & 10, 11 &\cellcolor{gray!60} 12, 13 & 14, 15 & 16, 17 & 18, 19 & \cellcolor{gray!60} 12, 13\tabularnewline
		\hline 		
	\end{tabular}
	\caption{Generated Mappings for the features of the Aircraft FM}
	\label{tab:Mappings}
\end{table*}
%}

Line~\ref{al:getMandRoot} uses auxiliary function \texttt{findMandAndRoot} to determine the root feature and all mandatory child features of the original input FM that is now stored in the input file \texttt{constr}. 
Function \texttt{findMandAndRoot} returns a tuple \texttt{(root, mandatories)}, where the first entry denotes the column that represents the root feature of the FM and the second entry contains a mapping for all mandatory children to their parent. 
%generate a mapping from old to new values and vice versa. To do so it first  determines the root feature and all mandatory child features. 
Because the constraints imposed by the FM are only available in CNF \texttt{findMandRoot} uses a SAT solver to perform the Root and Mandatory Child test as they are explained next\footnote{The representation of FMs in CNF has been extensively documented, please refer to \cite{Benavides2010}.}.   

\begin{definition}
Root test: Let \texttt{constraints} denote the constraints describing the FM in CNF and \texttt{f} be a feature of the FM, then it can be inferred that \texttt{f} is a root feature if $\texttt{constraints} \land \neg \texttt{f}$ cannot be satisfied.\footnote{Note that we return the first feature for which the Root test is satisfied.}
\end{definition}

\begin{definition}
Mandatory Child test: Let \texttt{constraints} denote the constraints describing the FM in CNF and \texttt{f$_1$} and \texttt{f$_2$} denote two features of the FM, then it can be inferred that \texttt{f$_2$} is a mandatory child of \texttt{f$_1$} if neither $\texttt{constraints} \land \neg \texttt{f$_1$} \land \texttt{f$_2$}$ nor $\texttt{constraints} \land \texttt{f$_1$} \land \neg \texttt{f$_2$}$ can be satisfied.%\footnote{It would also be valid to infer that \texttt{f$_1$} is a mandatory child of \texttt{f$_2$}.} 
\end{definition}

We can for instance infer that \texttt{Aircraft} is the root feature of the input FM, as the proposition \texttt{constraints} $\land \neg 0$ cannot be satisfied. 
We can also infer that \texttt{Rust} is a mandatory child of \texttt{Metal} because neither \texttt{constraints} $\land \neg 18 \land 26$ nor \texttt{constraints} $\land 18 \land \neg 26$ can be satisfied.

Next auxiliary function \texttt{generateMappings} computes two mappings, one from old to new values \texttt{oldToNew} and one vice versa \texttt{newToOld}, in the following way:

\begin{itemize}
	\item The root feature with original values \texttt{c$_1$} and \texttt{c$_2$} $=$ \texttt{c$_1$}+1 is mapped to the values \texttt{c$_1$'}$=2*(n-m)$ and \texttt{c$_2$'}$=2*(n-m)+1$. Note here that the root feature is mapped to values which will not be considered by the annealing procedure. 
	\item All features that are neither the root feature nor mandatory child features are assigned unique values from the range $0$ to $2 * (n-m) -1$.
	\item All features that are mandatory child features are mapped to the same values as their mandatory parent. 
\end{itemize}

The mapping of new to old values \texttt{newToOld} maps each new value to the corresponding old value.

Consider Table~\ref{tab:Mappings}, which shows the old and new values that represent the features of our running example.
We have 4 features that are neglectible according to the reduction rules, feature \texttt{Aircraft} because of Reduction Rule 1, and features \texttt{Wing, Material} and \texttt{Rust} because of Reduction Rule 2. Therefore, the root feature \texttt{Aircraft} is mapped to $c_1=2*(n-m)=2*(14-4)=20$ and $c_2=c_1+1=21$. Next the features that are needed to compute the tCA are assigned values from $0$ to $2*(n-m)-1=19$. Finally all mandatory child features are assigned the same values as their parent features, these assignments are accordingly highlighted in Table~\ref{tab:Mappings}. 
For instance the mapping \texttt{oldToNew(0)} yields value 20, and the mapping \texttt{newToOld(12)} yields value 18. 
Note that the mapping \texttt{oldToNew} is not bijective, meaning that two values may map to the same new value. We chose to model the inverse mapping \texttt{newToOld} in a way that the new value maps to the old value of the mandatory parent and not its mandatory child. For instance \texttt{newToOld(12)} maps to 18 that is used to represent the selection of feature \texttt{Metal} rather than to 26 that is used to represent the selection of feature \texttt{Rust}.

Next, Line~\ref{al:redFeatures} in Algorithm~\ref{al:overall} reduces the number of features. For our running example the number features is 10, hence the valid values per column now range from 0 to 19 rather than from 0 to 27.  
Subsequently Line~\ref{al:endAdapt} adapts the input constraints by replacing each old value by its new value. For instance the constraint that the optional child \texttt{Engine} implies its parent feature \texttt{Aircraft} that is denoted by $\neg 4 \lor 0$ is replaced by $ \neg \texttt{oldToNew(4)} \lor \texttt{oldToNew(0)}$, i.e. $\neg 0 \lor 20$.

\begin{sloppypar}
Subsequently Line~\ref{al:anneal} calls CASA's simulated annealing procedure with the reduced inputs \texttt{features'} and \texttt{properties'}. As a consequence the generated tCA does not 
adhere to the original constraints moreover there are no columns representing the root nor the mandatory features. Hence, we need to alter \texttt{tCA} in two ways, before it can be returned by CASA: 
\begin{itemize}
	\item We need to replace each value \texttt{v} in the generated array by its old value, i.e. \texttt{newToOld}.
	\item We need to expand \texttt{tCA}, i.e. add columns for the root feature and all mandatory child features. The root feature needs, by definition, to be selected in every row of the generated tCA and the mandatory child features are selected whenever their corresponding parent is included. 
\end{itemize}

\begin{figure}
\centering{}\includegraphics[scale=0.26]{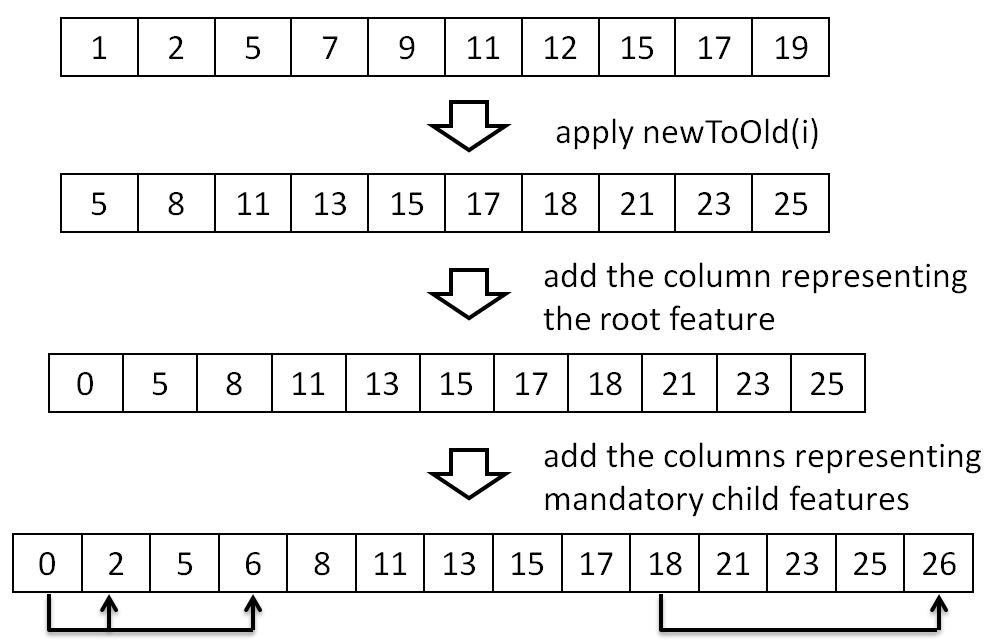}\caption{Example of expanding a row in the tCA}
\label{fig:axpand} 
\end{figure}

These adaptations of the generated tCA are performed by auxiliary function \texttt{expand} called in Line~\ref{al:expand}. 
Consider Figure~\ref{fig:axpand}. It depicts how a single row in the generated tCA is expanded. 
At first the generated tCA contains the row [1, 2, 5, 7, 9, 11, 12, 15, 17, 19], which is then adapted by substituting each value in this row by its old value, i.e. [\texttt{newToOld}(1), \texttt{newToOld}(2), \texttt{newToOld}(5), \texttt{newToOld}(7), \texttt{newToOld}(9), \texttt{newToOld}(11), \texttt{newToOld}(12), \texttt{newToOld}(15), \texttt{newToOld}(17), \texttt{newToOld}(19)]=[5, 8, 11, 13, 15, 17, 18, 21, 23, 25]. Next the rows of the tCA are expanded by the selection of the root feature, i.e. [0, 5, 8, 11, 13, 15, 17, 18, 21, 23, 25]. Next the tCA is expanded by the proper selection and deselection of the mandatory child features, i.e. features \texttt{Wing} and \texttt{Material} are added to the row (denoted by values 2 and 6) because their parent \texttt{Aircraft} is selected (denoted by value 0). Moreover feature \texttt{Rust} is selected (denoted by value 26) because its parent \texttt{Metal} is selected (denoted by value 18), which leaves us with the row [0, 2, 5, 6, 8, 11, 13, 15, 17, 18, 21, 23, 25, 26]. Finally Line~\ref{al:return} returns the generated tCA.  
\end{sloppypar}

%% file: Evaluation.tex
\section{Evaluation}
%This section first presents the adaptations we performed on CASA  and then proceeds by outlining the results concerning our runtime and tCA size analysis.

We evaluated the adapted Version of CASA on a Windows 7 System running at 3.2Ghz on an Intel Core i5, and with RAM of 8GB. Both the original CASA and our extension with the two reduction rules were initialized to use an initial temperature of 0.5, a cooling of -0.0001\% because this setting showed the best results in \cite{GCD11}. Moreover we set the strength of the tCA to generatoe to t=3. 

%\subsection{Statistical Analysis of the Obtained Results}
We evaluated the effect of the presented reduction rules by comparing the runtimes and array sizes of the adapted version of CASA to the one of the original version. To do so we used 133 publicly available FMs from the SPLOT website \cite{SPLOT} and run each algorithm a hundred times per model. The number of features of these 133 models ranges from 9 to 61, where the number of neglectible features is between 1 and 23.

The original version of CASA needs to consider between 672 and 287920 t-sets for the models of our evaluation. Figure~\ref{fig:tsetReduction} shows a histogram that summarizes the percentage of reduction in the number of t-sets when the reduction rules are applied. It can be seen that for the majority of our test cases more than 50\% of the t-sets can be ignored during the generation of the tCA.

\begin{figure}
\centering{}\includegraphics[scale=0.5]{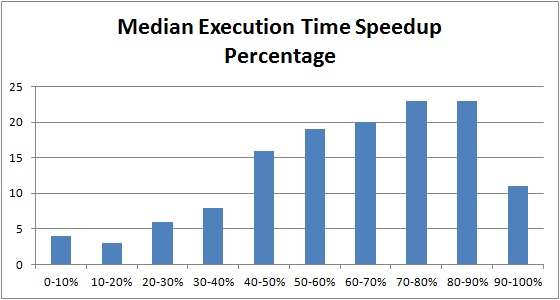}\caption{Reduction in the median execution times}
\label{fig:executionTimeRedution} 
\end{figure}

Next we considered the runtimes of both algorithms. The median execution time of the original CASA is between 375ms and 13.5 minutes. When the presented reduction rules are applied, the median execution time shows an average speedup of 61.8\% for the models of our evaluation. 
Figure~\ref{fig:executionTimeRedution} shows a histogram that summarizes the speedups of the median execution time in percentage for the 133 models used.

To ensure that the observed improvements of our timing analysis are not due to chance, we performed a thorough statistical analysis of the presented results. First we used the Shapiro-Wilk test to determine whether the recorded execution times stem from a normal distributed population \cite{DGMH11}. The highest p-value that we observed for the execution times of the adapted version of CASA is 0.0164\%, meaning that we can infer that the recorded execution times are not normal distributed. 
As a result, we performed the Wilcoxon Rank Sum test on each model to evaluate whether the adapted version of CASA showed indeed an improvement in the median execution time. For each of these tests we used the following null and alternative hypothesis:

\begin{figure}[h]
\centering{}\includegraphics[scale=0.5]{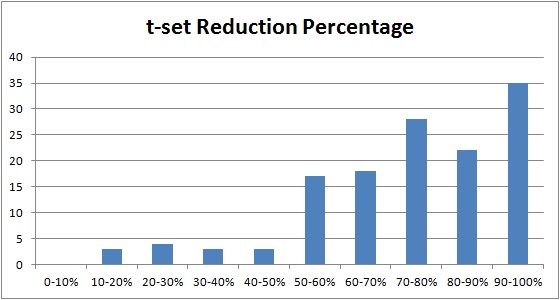}\caption{Reduction in the number of t-sets}
\label{fig:tsetReduction} 
\end{figure}

\begin{itemize}
	\item H$_0$: median$_{adapted}$ = median$_{original}$ \newline
	the median difference between the individual execution times is zero, i.e. there is no difference in the recorded runtimes.
	\item H$_1$: median$_{adapted}$ < median$_{original}$ \newline
	the median difference between the individual execution times is smaller zero, i.e. the execution time of the adapted version is smaller.
\end{itemize}

For 130 out of the 133 models we recorded p-values between 2.2 x 10$^{-16}$ and 4.84 x 10$^{-2}$, meaning that we can reject the null hypothesis in favor of the alternative hypothesis, i.e. accept that the presented reduction rules indeed show a statistically significant improvement of the observed runtimes. For the remaining 3 models the observed p-values where greater than 0.4, hence the null hypotheses could not be rejected. 
Two of these three models have very few features that could be eliminated (one feature and two features) during the generation of the tCA, this being the reason why our reduction rules show no significant improvement to the recorded execution times. For the third model we were able to neglect 50\% of the t-sets (the model contains 30 features, where 6 are reduceable), but still the adapted version of CASA showed a median execution time of 50014.5ms and the original version one of 49117ms. 

Lastly we used the Wilcoxon Signed Ranks Test\footnote{Again the Shapiro-Wilk test was used first to ensure that the examined data does not come from a normal distribution.} on the median execution time of the 133 models, using the same H$_0$ and H$_1$ as for the Wilcoxon Rank Sum Test. We observed a p-value of 2.20 x 10$^{-16}$ with R+=8721 and R-=190. Hence we can reject H$_0$ in favor of H$_1$ with a level of significance $\alpha = 0.01$, which means that we are able to infer that the presented reduction rules lead in general to shorter execution times. 

Lastly we also examined the size of the generated tCAs. For the majority of our test cases, i.e. 125 FMs, there is no difference in the median size of the generated arrays. For 6 FMs we produced tCAs, where the median size is between 3.75\% and 5.88\% smaller. For two models the generated tCAs, where 1.75\% and 4.44\% larger than the one produced by the original CASA version. Again we performed the Wilcoxon Rank Sum test for each of the evaluated FMs, where we could conclude for 17 models that the application of the presented reduction rules leads to potentially smaller arrays and for 2 models the original version of CASA tends to perform better. Finally the Wilcoxon signed ranks test on the median size of the tCAs over all models shows with a p-value of 0.1377\% that we cannot infer that the presented reduction rules lead to smaller tCAs sizes in general. All the data and statistical analysis are available online\footnote{URL omitted to presume anonymity}.

Overall our evaluation shows that even though we could not generate smaller tCAs, we were able to find them significantly faster than with the original version of CASA. 
In other words, we were able to speedup CASA without weakening the generated tCAs.  

%% file: RelatedWork.tex
\section{Related Work}
This section gives an overview of the existing related work in the context of SPLE testing, combinatorial interaction testing, existing algorithms to generate tCAs for SPLs and examples of search based techniques that have been applied to problems in SPLE. 

Different approaches have been proposed to overcome some of the difficulties of testing SPLs, for instance 
\textit{ScenTED} (SCENario based TEst case Derivation) is an approach to generate test case scenarios from UML models that contain variability information \cite{RRKP06}. Another example is Incremental testing, which aims to automatically adapt the test suite for a certain product by using knowledge about the commonalities and differences among the products of the SPL \cite{UKB10}. 

A survey about SPL testing by Engstr{\"o}m et. al shows that the focus of current research lies amongst other things on test organization, test management and high-level test case derivation \cite{ER11}. They mention CIT as one of the approaches for test management. 
Another survey by Lee et. al identifies for instance the test case creation and selection as well as the test execution for absent variants as research areas for SPL testing, where most of them are only marginally dealt with yet \cite{LKL12}. Lee et. al identify CIT as one of the approaches for the test case creation. 
To the best of our knowledge both surveys do not explicitly list any search based techniques that have been applied to SPL testing.

Changhai et. al performed a survey on CIT in general \cite{NL11}. They found that there exist four different research fields for generating tCAs: greedy
algorithms, heuristic search algorithms, mathematic methods, and random methods. 
Hill climbing, great flood, tabu search, simulated annealing, genetic algorithms and ant colony algorithms are listed as search based techniques that have been applied to generate tCAs. 
Changhai et. al point out that most test case generation algorithms cannot deal with constraints and simply ignore them.  
Simulated annealing is mentioned as one of the techniques that where applied to generate tCAs in the presence of constraints. 
%The survey does not explicitly mention approaches for generating tCAs in the context of SPLE. 

Let us now focus on examples of tCA generation algorithms for SPLs. 
ICPL is a recursive acronym and stands for "ICPL Covering array generation algorithm for Product Lines" \cite{GCD11}, it is a greedy approach to generate tCAs. Basically ICPL is an adaptation of Chv{\'a}tal's greedy algorithm to solve the set-cover problem\footnote{Note here that the generation of tCAs is an instance of the set-cover problem \cite{VaMosPaper13, GCD11}.}. 
As addition to their earlier work published in \cite{JHF11}, they performed several adaptations to improve ICPLs execution time. They for instance parallelize the data independent procedures of their algorithm. Moreover, they use the fact that a (t-1)-wise CA is always a subset of a t-wise CA \cite{FCP09}, meaning they can build up their tCA in a recursive manner. 
Johansen et. al compared their algorithm to CASA and MoSo-PoLiTe, where they claim that ICPL is faster than the other two algorithms and tends to generate smaller tCAs. 
Note that Haslinger et. al applied the proposed reduction rules to ICPL and could record speedups of up to 88\% \cite{VaMosPaper13}.

Oster et. al present MoSo-PoLiTe the Model-based Software Product Line testing framework \cite{OMR10}. MoSo-PoLiTe has been evaluated to generate 2-wise CAs for SPLs, i.e. they generate CAs only for pairwise testing. It is part of their future work to also evaluate MoSo-PoLiTe for 3-wise and 4-wise CAs. The overall procedure of MoSo-PoLiTe first flattens the input FM and then converts it into a \textit{Constraint Solving Problem (CSP)}. Next they apply forward checking on the CSP that corresponds to the input FM to generate a 2-wise CA.   
In contrast with our work of extending CASA, neither ICPL nor MoSo-PoLiTe rely on search based algorithms. 

The works of Lopez-Herrejon et. al \cite{LGBSE12} and Guo et. al \cite{GWWLW11} are examples of approaches where search based techniques have been applied to SPLE problems. Lopez-Herrejon et. al use in \cite{LGBSE12} a genetic algorithm to reverse engineer feature models from a set of feature sets. Guo et. al designed in \cite{GWWLW11} a Genetic Algorithm to find an optimal feature selection for given requirements when resource constraints are present. 

%% file: ConclusionsFutureWork.tex
\section{Conclusions and Future Work}
%In this research-in-progress paper we propose a set of reduction rules to reduce the number of t-sets that have to be covered by any t-wise CA generation algorithm, without weakening the strength of the generated arrays. We evaluated these rules by adapting ICPL \cite{JHF12}, where we recorded speedups of up to 88\% in median execution time. 
%
%Our reduction rules are not only applicable to ICPL, but to any tCA generation algorithm. Hence, we plan to evaluate our rules also on CASA \cite{GCD11} and MoSo-PoLiTe \cite{OMR10}. We also plan to explore Search Based algorithms such as that presented by Ferrer et al. \cite{FKCA12}.
%
%Additionally we have one more reduction rule in mind, which claims that any t-set that contains a feature \texttt{f} and one of its descendants does not need to be covered. It is part of our future work to formally proof this claim and to extend our current evaluation by also considering this rule.

By applying Haslinger et. al's reduction rules \cite{VaMosPaper13} to CASA --a simulated annealing algorithm, we provide further proof that these rules lead to significantly shorter runtimes when they are applied to tCA generation algorithms. 

Moreover there exists a third reduction rule, which states that any t-set that contains a feature \texttt{f} and one or more of its descendants \texttt{d} does not need to be considered during the computation of a tCA. This third rule is still not formally proven, especially not in the presence of Cross Tree Constraints. We need to assess if this additional reduction rule could be applied to CASA, as it is designed in a way that does not permit changes in the search without the need to restructure various parts of the source code. 

Of course apart from the runtime also the size of the generated tCAs plays an important role, to which the reduction rules do not seem to have any effect. 
Because the generation of tCAs in the context of SPLE is NP-complete we strongly believe that the their computation is extremely well suited for SBSE techniques. 
We therefore plan on investigating in other search based techniques to generate tCAs, such as genetic algorithms or ant colony algorithms. A good starting point would be the work by proposed by Ferrer et. al \cite{FKCA12}. 

%Both CASA and ICPL heavily rely on SAT solvers. It is part of our future work to assess whether a feature model reasoner (i.e. an algorithm that operates on the tree-structure of the FM to validate whether certain product configuration is valid) might show better performance. 

%% file: gecco.bbl
\begin{thebibliography}{10}

\bibitem{SPLOT}
{Software Product Line Online Tools(SPLOT)}, Accesed July 2011.
\newblock http://www.splot-research.org/.

\bibitem{BSR04}
D.~Batory, J.~N. Sarvela, and A.~Rauschmayer.
\newblock {Scaling Step-Wise Refinement}.
\newblock {\em IEEE TSE}, 30(6), 2004.

\bibitem{Benavides2010}
D.~Benavides, S.~Segura, and A.~Ruiz-Cort�s.
\newblock Automated analysis of feature models 20 years later: A literature
  review.
\newblock {\em Information Systems}, In Press, Corrected Proof:--, 2010.

\bibitem{Cohen08}
M.~Cohen, M.~Dwyer, and J.~Shi.
\newblock Constructing interaction test suites for highly-configurable systems
  in the presence of constraints: A greedy approach.
\newblock {\em Software Engineering, IEEE Transactions on}, 34(5):633 --650,
  sept.-oct. 2008.

\bibitem{CE00}
K.~Czarnecki and U.~Eisenecker.
\newblock {\em {Generative Programming: Methods, Tools, and Applications}}.
\newblock Addison-Wesley, 2000.

\bibitem{DBLP:conf/splc/2012-1}
E.~S. de~Almeida, C.~Schwanninger, and D.~Benavides, editors.
\newblock {\em 16th International Software Product Line Conference, SPLC '12,
  Salvador, Brazil - September 2-7, 2012, Volume 1}. ACM, 2012.

\bibitem{DGMH11}
J.~Derrac, S.~Garc\'{\i}a, D.~Molina, and F.~Herrera.
\newblock A practical tutorial on the use of nonparametric statistical tests as
  a methodology for comparing evolutionary and swarm intelligence algorithms.
\newblock {\em Swarm and Evolutionary Computation}, 1(1):3--18, 2011.

\bibitem{ER11}
E.~Engstr{\"o}m and P.~Runeson.
\newblock Software product line testing - a systematic mapping study.
\newblock {\em Information {\&} Software Technology}, 53(1):2--13, 2011.

\bibitem{FKCA12}
J.~Ferrer, P.~M. Kruse, J.~F. Chicano, and E.~Alba.
\newblock Evolutionary algorithm for prioritized pairwise test data generation.
\newblock In T.~Soule and J.~H. Moore, editors, {\em GECCO}, pages 1213--1220.
  ACM, 2012.

\bibitem{FCP09}
S.~Fouch{\'e}, M.~B. Cohen, and A.~A. Porter.
\newblock Incremental covering array failure characterization in large
  configuration spaces.
\newblock In G.~Rothermel and L.~K. Dillon, editors, {\em ISSTA}, pages
  177--188. ACM, 2009.

\bibitem{GCD11}
B.~J. Garvin, M.~B. Cohen, and M.~B. Dwyer.
\newblock Evaluating improvements to a meta-heuristic search for constrained
  interaction testing.
\newblock {\em Empirical Software Engineering}, 16(1):61--102, 2011.

\bibitem{GWWLW11}
J.~Guo, J.~White, G.~Wang, J.~Li, and Y.~Wang.
\newblock A genetic algorithm for optimized feature selection with resource
  constraints in software product lines.
\newblock {\em Journal of Systems and Software}, 84(12):2208--2221, 2011.

\bibitem{FasePaper13}
E.~N. Haslinger, R.~E. Lopez-Herrejon, and A.~Egyed.
\newblock On extracting feature models from sets of valid feature combinations.
\newblock In {\em to appear at the 16th International Conference on Fundamental
  Approaches to Software Engineering (FASE)}, 2013.

\bibitem{VaMosPaper13}
E.~N. Haslinger, R.~E. Lopez-Herrejon, and A.~Egyed.
\newblock Using feature model knowledge to speed up the generation of covering
  arrays.
\newblock In {\em to appear at the Seventh International Workshop on
  Variability Modelling of Software-intensive Systems (VAMOS)}, 2013.

\bibitem{JHF11}
M.~F. Johansen, {\O}.~Haugen, and F.~Fleurey.
\newblock Properties of realistic feature models make combinatorial testing of
  product lines feasible.
\newblock In J.~Whittle, T.~Clark, and T.~K{\"u}hne, editors, {\em MoDELS},
  volume 6981 of {\em Lecture Notes in Computer Science}, pages 638--652.
  Springer, 2011.

\bibitem{JHF12}
M.~F. Johansen, {\O}.~Haugen, and F.~Fleurey.
\newblock An algorithm for generating t-wise covering arrays from large feature
  models.
\newblock In de~Almeida et~al. \cite{DBLP:conf/splc/2012-1}, pages 46--55.

\bibitem{KCH+90}
K.~Kang, S.~Cohen, J.~Hess, W.~Novak, and A.~Peterson.
\newblock {Feature-Oriented Domain Analysis (FODA) Feasibility Study}.
\newblock Technical Report CMU/SEI-90-TR-21, Software Engineering Institute,
  Carnegie Mellon University, 1990.

\bibitem{KWG04}
D.~R. Kuhn, D.~R. Wallace, and A.~M. Gallo, Jr.
\newblock Software fault interactions and implications for software testing.
\newblock {\em IEEE Trans. Softw. Eng.}, 30(6):418--421, June 2004.

\bibitem{LKL12}
J.~Lee, S.~Kang, and D.~Lee.
\newblock A survey on software product line testing.
\newblock In de~Almeida et~al. \cite{DBLP:conf/splc/2012-1}, pages 31--40.

\bibitem{LGBSE12}
R.~E. Lopez-Herrejon, J.~Galindo, D.~Benavides, S.~Segura, and A.~Egyed.
\newblock Reverse engineering feature models with evolutionary algorithms: An
  exploratory study.
\newblock In G.~Fraser and J.~T. de~Souza, editors, {\em SSBSE}, volume 7515 of
  {\em Lecture Notes in Computer Science}, pages 168--182. Springer, 2012.

\bibitem{NL11}
C.~Nie and H.~Leung.
\newblock A survey of combinatorial testing.
\newblock {\em ACM Comput. Surv.}, 43(2):11:1--11:29, Feb. 2011.

\bibitem{OMR10}
S.~Oster, F.~Markert, and P.~Ritter.
\newblock Automated incremental pairwise testing of software product lines.
\newblock In J.~Bosch and J.~Lee, editors, {\em SPLC}, volume 6287 of {\em
  Lecture Notes in Computer Science}, pages 196--210. Springer, 2010.

\bibitem{SPLE}
K.~Pohl, G.~Bockle, and F.~J. van~der Linden.
\newblock {\em Software {P}roduct {L}ine {E}ngineering: {F}oundations,
  {P}rinciples and {T}echniques}.
\newblock Springer, 2005.

\bibitem{PBL05}
K.~Pohl, G.~Bockle, and F.~J. van~der Linden.
\newblock {\em Software {P}roduct {L}ine {E}ngineering: {F}oundations,
  {P}rinciples and {T}echniques}.
\newblock Springer, 2005.

\bibitem{RRKP06}
A.~Reuys, S.~Reis, E.~Kamsties, and K.~Pohl.
\newblock The scented method for testing software product lines.
\newblock In T.~K{\"a}k{\"o}l{\"a} and J.~C. Due{\~n}as, editors, {\em Software
  Product Lines}, pages 479--520. Springer, 2006.

\bibitem{UKB10}
E.~Uzuncaova, S.~Khurshid, and D.~S. Batory.
\newblock Incremental test generation for software product lines.
\newblock {\em IEEE Trans. Software Eng.}, 36(3):309--322, 2010.

\bibitem{zave}
P.~Zave.
\newblock Faq sheet on feature interaction.
\newblock http://www.research.att.com/~pamela/faq.html.

\end{thebibliography}
